\documentstyle[12pt]{article}
\newcommand{\be}{\begin{equation}}
\newcommand{\ee}{\end{equation}}
\newcommand{\bea}{\begin{eqnarray}}
\newcommand{\eea}{\end{eqnarray}}
\newcommand{\bc}{\begin{center}}
\newcommand{\ec}{\end{center}}
\newcommand{\vac}{\mid 0\rangle}
\newcommand{\x}{{\vec{\rm x}}}
\newcommand{\ka}{{\vec{\rm k}}}
\newcommand{\y}{{\vec{\rm y}}}
\newcommand{\q}{{\vec{\rm q}}}
\newcommand{\p}{{\vec{\rm p}}}
\newcommand{\vs}{{\vec{\rm v}}}

\newcommand{\rv}{{\vec{\rm r}}}
\newcommand{\rr}{{\vec{\varrho }}}
\newcommand{\s}{{\vec{\rm s}}}
\newcommand{\lv}{{\vec{\rm l}}}
\newcommand{\jv}{\vec{J}}
\newcommand{\kap}{\vec{\kappa}}
\newcommand{\di}{\displaystyle \int }
\newcommand{\NM}{{N^{\dagger }_{in} }}
\newcommand{\TY}{{\Theta^{\dagger }_{in} }}
\newcommand{\N}{{N_{in} }}
\newcommand{\T}{{\Theta_{in} }}
\newcommand{\PP}{{\cal P}}
\newcommand{\Pop}{{\mbox{\bf P}} }
\newcommand{\Xop}{{\mbox{\bf X}} }
\newcommand%
{\E}%
[1]%
{E_2({\cal P},#1)}%
\newcommand%
{\I}%
[1]%
{{I^{{\cal P}(\pm)}_{#1}(q)}}%
\newcommand%
{\VF}%
[1]%
{\Phi^{\pm}_{{\cal P}q}(#1)}%
\newcommand%
{\vf}%
[2]%
{\phi^{\pm (#1)}_{{\cal P}q}(#2)_{\alpha\beta}}%
\newcommand%
{\F}%
[2]%
{{\cal F}^{\cal P}_{#1}(#2;\kappa )}%
\newcommand%
{\f}%
[1]%
{f^{\cal P}_{#1}(\sigma;\kappa )}%
\newcommand%
{\J}%
[1]%
{{\rm J}^{\cal P}_{#1}(\sigma)}%
\newcommand%
{\CC}%
[1]%
{{{\cal C}^{{\cal P}(\pm)}_{#1}(q)}}%
\newcommand%
{\DD}%
[1]%
{{{\cal D}^{{\cal P}(\pm)}_{#1}(q)}}%
\newcommand%
{\TT}%
[2]%
{{{\cal T}^{(#1)\pm}_{{\cal P}}(q;#2)}}%
\newcommand%
{\kaK}%
[1]%
{{\cal K}^{\cal P}_{#1}(\eta-\tau)}%
\newcommand%
{\D}%
[1]%
{D^{\cal P}_{#1}(\sigma)}%
\begin{document}

\begin{center}
{\Large \bf Dynamical mapping method in nonrelativistic models of
quantum field
theory }
\footnote{This work is partially supported by RFFR N 94-02-05204
and by grant "Universities of Russia", (St-Pb)
N 94-6.7-2057.}
\end{center}

\vspace{4mm}

\begin{center}
A.N.Vall, S.E.Korenblit, V.M.Leviant, A.B.Tanaev. \\
Irkitsk State University, 664003, Gagarin bl-d, 20, Irkutsk, Russia.
\footnote{E-mail VALL@physdep.irkutsk.su, KORENB@math.isu.rannet.ru }
\end{center}

\begin{abstract}
{\small The solutions of Heisenberg equations and two-particles
eigenvalue problems for nonrelativistic models of current-current
fermion interaction and $N, \Theta $ model are obtained in the frameworks
of dynamical mapping method. The equivalence of different types of
dynamical mapping is shown. The connection between renormalization procedure
and theory of selfadjoint extensions is elucidated. }
\end{abstract}
\begin{center}
{\bf 1. General consideration}
\end{center}

The main problem of QFT follows from the fact that any solutions of
Heisenberg equations (HE) are the operator distributions which products,
always appearing in that equations, are ill-defined.
\be
\left(i\partial_t-{\cal E}({\Pop})\right)\Psi_{\alpha}(\x ,t)=
\left[\Psi_{\alpha}(\x ,t)\,,{\rm H}_I\{\Psi\}\right ]=?\;\;
\mbox{for}\quad {\rm H}\{\Psi\}={\rm H}_0\{\Psi\}+{\rm H}_I\{\Psi\}
\label{HE}
\ee
So, the correct definition of field equations (and Hamiltonian itself)
implies some knowledge about qualitative properties of its solutions which
in their turn depend on the form of these equations by a very singular manner.
The usual way to go out from this closed circle is connected with
perturbation theory. It is based on the assumption that product of
Heisenberg fields (HF) may be defined as well as for the free
ones and solution of HE may be obtained by perturbation in the Fock space
of renormalized {\sl free} fields.
However, it is impossible on such a way to work with nonrenormalizable theory
and to understand the origin of the bound states.
We consider another possibility which is based on the idea of dynamical
mapping
and reduce the product of HF to the normal ordering for the product of the
{\sl physical fields}. It is originated from the works of R.Haag \cite{Haag},
O.Greenberg \cite{Greenb}, H.Umezawa \cite{Umz}, and L.D.Faddeev \cite{Fadd}
M.I.Shirokov \cite{Shir} (see also \cite{Fiv}).

In this approach the problem of making a sense for formal expression of HF:
\be
\Psi (\x,t)=e^{i{\rm H}(t-t_0)}\, \Psi (\x,t_0) \,e^{-i{\rm H}(t-t_0)}
\Longrightarrow {\cal F}^t \left[ \Psi (\x,t_0) \right],
\label{HF}
\ee
for Hamiltonian given as a functional
${\rm H}= {\rm H} \left[ \Psi (\x,t) \right]$,
is divided on two parts. The first one is the construction of the
following operator realization of the initial fields
$\Psi ({\x},t_0)= \Psi [\psi] $ via {\sl physical fields}
$\psi (\x ,t)\equiv \psi \left\{A_{\alpha }(\ka )\right\}$,
which, on the one hand, should be consistent with CCR (CAR) ($\alpha = 1,2$)
\bea
&& \{ \Psi_{\alpha }(\x ,t)\,, \,\Psi_{\beta }(\y ,t) \}=0=
\{\psi_{\alpha } (\x ,t)\,,\, \psi_{\beta }(\y ,t) \},
\nonumber \\
&&\{ \Psi_{\alpha }(\x ,t)\,, \,\Psi_{\beta }^{\dagger}(\y ,t) \}
= \delta_3(\x - \y )\;\delta_{\alpha \beta } =
\{\psi_{\alpha } (\x ,t)\,,\, \psi_{\beta }^{\dagger} (\y ,t) \},
\nonumber \\
&& \{A_{\alpha }(\ka )\,,\, A_{\beta }(\q )\}=0;\;\;\;
\{A_{\alpha }(\ka )\,,\, A_{\beta }^{\dagger }(\q )\}=\delta_3(\ka-\q )\,
\delta_{\alpha \beta },
\label{CAR}
\eea
and on the other hand leads to unique stable vacuum $\vac $ and one-particle
state $\mid 1\ka ,\alpha \rangle $ with definite spectrum $E(\ka)$:
\bea
&& H\mid 0 \rangle =V\,w_0 \vac ;\qquad  A_{\alpha }(\ka )\mid 0\rangle =0 ;
\qquad V\;\mbox{is space volume};
\label{0pST} \\
&& [H,A^{\dagger}_{\alpha }(\ka )]\mid 0 \rangle =
E(k) A^{\dagger}_{\alpha }(\ka )
\mid 0 \rangle=E(k) \mid 1\ka ,\alpha \rangle =
(H-V\,w_0) \mid 1\ka ,\alpha \rangle.
\label{1pST}
\eea
Moreover, let us suppose that for such operator realization the reduced (time-
independent) Hamiltonian for the definite moment $t=t_0$ does not contain
"fluctuation" terms \cite{Schweb} up to fourth order and looks like :
\bea
&& H\equiv H\{A\}=V\,w_0 +\hat{H}\{A\}=
:{\rm H}\left\{\Psi \left[\psi [A\right] \right\}:
= V\,w_0+\hat{H}_0\{A\}+\hat{H}_I\{A\};
\nonumber \\
&& \hat{H}\mid 0\rangle =\hat{H}_0\mid 0\rangle =\hat{H}_I\mid 0\rangle = 0;
\label{Hred} \\
&& \hat{H}_0\{A\} \stackrel{def}{\equiv} \di d^3k E(k)
A^{\dagger}_{\alpha }(\ka )A_{\alpha }(\ka ); \;\;\;
[\hat{H}_0, A^{\dagger}_{\alpha }(\ka )]\equiv
E(k)A_{\alpha }^{\dagger}(\ka );
\nonumber \\
&& \hat{H}_I\{A\}=\di d^3q \di d^3p\di d^3\kappa \di d^3l \;
\delta_3(\q +\p -\kap -\lv )\; K^{q+p}_{22}
\left(\frac{\q -\p}{2},\frac{\kap -\lv}{2}\right)\cdot
\nonumber \\
&& \cdot A^{\dagger}_{\alpha } (\kap )A^{\dagger}_{\beta }(\lv )
A_{\beta }(\p)\,A_{\alpha }(\q)\;+\sim\sum
(A^{\dagger})^m (A)^n;\qquad m,n\geq 3.
\nonumber \\
&& K^{\PP}_{22}(\rv ,\s )=K^{\PP}_{22}(-\rv ,-\s );
\qquad \stackrel{*}{K^{\PP}_{22}}(\s ,\rv ) =K^{\PP}_{22}(\rv ,\s );
\nonumber
\eea
The general consideration of the existence of the operator realizations of
such kind is a subject of our another work. They always exist for the Lee
models considered below.

This work will be concentrated on the second part of the problem which is the
construction of the corresponding dynamical mapping (Haag expansion)
${\cal F}^t \left[ \Psi (\x,t_0) \right]$ as a series of normal ordered
products of {\sl physical fields} $ \Psi (\x,t_0) $ or $\psi (\x),\;\;A(\ka)$:
\bea
 e^{iH(t-t_0)} A_{\alpha }(\ka ) e^{-iH(t-t_0)} \equiv e^{-iE(k)(t-t_0)}
a_{\alpha }(\ka ,t)=e^{-iE(k)(t-t_0)}\;{\cal A}^t_{t_0}\left[A_{\alpha}(\ka )
\right],
\label{dm1} \\
 a_{\alpha}(\ka,t) = A_{\alpha }(\ka ) +\di d^3l \di d^3p\,
A^{\dagger}_{\beta }(\lv +\p ) A_{\beta } (\ka +\p ) A_{\alpha }(\lv )\,
F^{(1)}_A (t;\p \mid \lv ,\ka ) +\ldots
\label{dm2} \\
 \Rightarrow \;\;(\mbox{for}\;m=n)\;\;\Rightarrow
{\cal A}^t_{t_0}\left[A_{\alpha}(\ka)\right] =A_{\alpha}(\ka)+
\sum\limits^{\infty}_{n=1}
\di d^3l\prod\limits^n_{j=1}\left\{\di d^3q_j\di d^3p_j \right\}\cdot
\nonumber \\
 \cdot \prod\limits^n_{j=1}\left\{A^{\dagger}_{\beta_j}(\q_j) \right\}
\prod\limits^1_{j=n}\left\{A_{\beta_j}(\p_j) \right\}A_{\alpha}(\lv)\,
Y^{(n)}_A(t;\ka;\{\q_j\}^n_1\,|\{\p_j\}^1_n;\,\lv),
\label{dm3}
\eea
and the usage of these coefficient functions for eigenvalue problem.
The condition $m=n$ in the last eq.(\ref{dm3}) means the absence of
"fluctuation terms" (with $m\neq n$) in reduced Hamiltonian (\ref{Hred}),
which commutes with particle number operator for that case.

From the expressions (\ref{dm1}),(\ref{dm2})  it's follows that vacuum and
one -particle states remain stable for all $t$:
\bea
&& a_{\alpha }(\ka ,t) \mid 0 \rangle \Longrightarrow A_{\alpha }(\ka )
\mid 0 \rangle \equiv 0; \quad
  \mid 1\ka ,\alpha \rangle = a^{\dagger}_{\alpha }(\ka ,t) \mid 0 \rangle
\Longrightarrow A^{\dagger}_{\alpha }(\ka )\mid 0 \rangle;
\nonumber \\
&&  \left[H,\,a^{\dagger}_{\alpha }(\ka ,t) \right]\mid 0\rangle =
E(k)\, A^{\dagger}_{\alpha }(\ka )\mid 0 \rangle,
\label{Aegnv}
\eea
what allows one to {\sl define} the normal ordering directly for HF and the
normal ordered Hamiltonian (\ref{Hred}) now correctly defines the nonlinear
terms in reduced HE (\ref{HE}):
\bea
&& \left(i\partial_t -E({\Pop}) \right) \Psi_{\alpha} (\x ,t)=
\left[ \Psi_{\alpha} (\x ,t)\,,\hat{H}_I\{\Psi\}\right ].
\label{Geq?} \\
&& i\partial_t a_{\alpha}(\ka,t)=\left[ a_{\alpha}(\ka,t),\,\hat{H}_I\{a\}
\right] \Rightarrow \di d^3l\,Q_{(a)}(\ka,\lv;t)\,a_{\alpha}(\lv,t);
\label{Geq??} \\
&& Q_{(a)}(\ka,\lv;t)=\di d^3q\di d^3p\,a^{\dagger}_{\beta}(\q,t)\,
a_{\beta}(\p,t)\,
e^{it[E(k)+E(q)-E(p)-E(l)]}\cdot
\nonumber \\
&& \delta_3(\ka+\q-\p-\lv)\; \frac{2}{i}\;
K^{\lv+\p}_{22}\left(\frac{\lv-\p}{2},\frac{\ka-\q}{2}\right)=
\label{Q-K} \\
&& =\di d^3\varrho\,
a^{\dagger}_{\beta}(\lv+\rr,t)\,a_{\beta}(\ka+\rr,t)\,
e^{it[E(\lv+\rr)+E(\ka)-E(\ka+\rr)-E(\lv)]}\cdot
\nonumber \\
&& \cdot \frac{2}{i}\; K^{\ka+\lv+\rr}_{22}\left(\frac{\lv-\ka-\rr}{2},
\frac{\ka-\lv-\rr}{2} \right).
\nonumber
\eea
The case (\ref{dm3}) means, moreover, the stability and absence of any
polarization not only for vacuum and one-particle states but also for
arbitrary $N$-particle ones. So, for arbitrary $N$ one can reduce the product:
\be
\langle 0\mid\,\prod \limits^N_{i=1} a_{\alpha}(\ka_i,t)
\stackrel{to}{\longrightarrow}\sim
\langle 0\mid\,\sum \,\prod \limits^N_{i=1} \di d^3s_i \,A_{\alpha}(\s_i).
\label{vG-F}
\ee
However, if fluctuation terms appear with min $(m,n) \geq N_0$, then such
reduction (\ref{vG-F}) is possible only for $N < N_0$.

There exist two essentially different choices for the initial moment $t_0$
leading to corresponding different choices of physical fields: \\
\begin{tabular}{c|c|c}
$t_0\rightarrow -\infty$, (Greenberg, Umezawa) & &
$t_0=0$, (Faddeev, Shirokov)\\
nonoperator initial condition & & operator initial condition \\
${\rm w}\lim\limits_{t\to -\infty}\langle f_{in}\mid \Psi (\x,t)-
\psi_{in}(\x,t)\mid i_{in}\rangle =0 $ & & $ {\rm s}\lim\limits_{t\to 0}
\Psi (\x,t)=\Psi\left[\psi(\x.0)\right]$  \\
$ \left\{\psi_{in}[A_{in}]\right\}\rightarrow $ incomplete Fock space & &
$ \left\{\psi [A]\right\}\rightarrow $ complete Fock space \\
new fields $V_{in}$ for every bound state & & no any new fields for bound
states \\
$ \hat{H}\stackrel{weak}{=}\hat{H}_0\{A_{in}\}+\hat{H}_0\{V_{in}\}+
\ldots $ & & $ \hat{H}=\hat{H}_0\{A\}+\hat{H}_I\{A\}$ \\
\end{tabular} \\
The second choice $t_0=0$ is used here. It seems more economical, and both
bound and scattering eigenstates look equal in rights for that choice.

One can check by direct substitution that solutions of both scattering and
bound state two-particles eigenvalue problems
\bea
&& \hat{H}\mid R^{\pm}_{\alpha \beta} (\PP,\q ) \rangle  =
\E{q}\mid R^{\pm}_{\alpha \beta} (\PP ,\q ) \rangle;
\;\;\;\E{q}\equiv E(\frac{\PP}{2}+\q ) + E(\frac{\PP}{2}-\q );
\nonumber \\
&& \hat{H}\mid {\rm B}^{\PP}_{\alpha \beta} \rangle  =
M_2(\PP)\mid {\rm B}^{\PP}_{\alpha \beta} \rangle;
\label{egvlpr}
\eea
{\sl exist} in the Fock eigenspace of kinetic part of reduced Hamiltonian
(\ref{Hred})
\be
\hat{H}_0\{A\}\, A^{\dagger}_{\alpha_1 }(\ka_1) \cdot \cdot \cdot
A^{\dagger}_{\alpha_n } (\ka_n) \mid 0 \rangle =\left (\displaystyle
\sum_{j=1}^n E(k_j)  \right ) A^{\dagger}_{\alpha_1 }(\ka_1)\cdot \cdot \cdot
A^{\dagger}_{\alpha_n }(\ka_n) \mid 0 \rangle,
\label{HOSpc}
\ee
in the following form:
\bea
&& \mid R^{\pm}_{\alpha \beta} (\PP ,\q ) \rangle = \di d^3 \kap
\mid R^0_{\alpha \beta} (\PP ,\kap ) \rangle \Phi^{\pm}_{\PP q}(\kap);\;\;
\mid {\rm B}^{\PP}_{\alpha \beta}\rangle = \di d^3 \kap
\mid R^0_{\alpha \beta} (\PP ,\kap ) \rangle B^{\PP}(\kap);
\nonumber \\
&& \mid R^0_{\alpha \beta} (\PP,\q ) \rangle \equiv A^{\dagger}_{\alpha }
(\frac{\PP}{2}+\q ) A^{\dagger}_{\beta }(\frac{\PP}{2}-\q ) \mid 0 \rangle,
\label{EgnSoln}
\eea
where corresponding wavefunctions satisfy to the usual Lippman-Schwinger
equations:
\bea
&& \Phi^{\pm}_{\PP q} (\kap ) = \delta_3(\kap -\q ) +
\frac{1}{E_2(\PP ,q) -E_2(\PP ,\kappa) \pm i\delta}\cdot
\di d^3r\; \Phi^{\pm}_{\PP q}(\rv ) \; 2\,K^{\PP}_{22}(\rv ,\kap ),
\nonumber \\
&& B^{\PP}(\kap ) = \frac{1}{M_2(\PP) -E_2(\PP ,\kappa) }\cdot
\di d^3r\; B^{\PP}(\rv ) \; 2\,K^{\PP}_{22}(\rv ,\kap ).
\label{LpmnSchwgr}
\eea
In its turn, at $m,n \geq 3$ for the first coefficient function of
(\ref{dm2})
\be
Y^{(1)}_A(t;\ka;\q\mid\p;\lv)\equiv\delta_3(\ka+\q-\p-\lv)\;
F^{(1)}_A(t;\p-\ka\mid\lv,\ka)
\label{Y1F}
\ee
from reduced HE (\ref{Geq??}) follows an integral equation with the same
kernel:
\bea
&&  F^{(1)}_A\left(t;-\kap -\q\mid\frac{\PP}{2}+\kap ,\frac{\PP}{2}+\q \right)
\equiv F^{(1)}_A\left(t;\kap+\q\mid\frac{\PP}{2}-\kap ,\frac{\PP}{2}-\q\right)=
\nonumber \\
&& = \di \limits^{t}_0 d\eta \,e^{i\eta [E_2(\PP ,q) - E_2(\PP ,\kappa) ]}
\left [ \frac{2}{i}\, K^{\PP}_{22}(\kap ,\q )
+ \di d^3r \,e^{i\eta [E_2(\PP ,\kappa) - E_2(\PP ,r) ]}  \cdot \right.
\nonumber \\
&& \left. \cdot \frac{2}{i}K^{\PP}_{22}(\rv ,\q )\,
F^{(1)}_A\left(\eta ;\kap +\rv\mid \frac{\PP}{2}-\kap ,\frac{\PP}{2}-\rv \right)
\right].
\label{Feq}
\eea
It contains all information about two-particle sector, directly determining
the scattering wave function for
$E_2(\PP ,q)\Rightarrow E_2(\PP ,q) \pm i\delta $:
\be
\Phi^{\pm}_{\PP q} (\kap ) = \delta_3(\kap -\q )+ F^{(1)*}_A
\left(t=\mp \infty;-\kap -\q \mid \frac{\PP}{2}+\kap ,\frac{\PP}{2}+\q \right),
\label{Scatt_F}
\ee
where the simply derived expression for scattering state was used:
\bea
&& \mid R^{\pm}_{\alpha \beta} (\PP,\q ) \rangle  =
\mid R^{0}_{\alpha \beta} (\PP ,\q ) \rangle +\di d^3 \kappa\,2i\,
K^{\PP}_{22}(\q ,\kap )\cdot
\nonumber \\
&& \cdot \di \limits^{\mp \infty}_0 dt\;
e^{-it\left(\E{q}-\E{\kappa}\pm i\delta \right)}\,
a^{\dagger}_{\alpha}(\frac{\PP}{2}+\kap,t)\,
a^{\dagger}_{\beta}(\frac{\PP}{2}-\kap,t)\,\mid 0 \rangle,
\label{Prdst2}
\eea
which follows directly from (\ref{dm1}) and definition of scattering state:
\be
\mid R^{\pm}_{\alpha \beta} (\PP,\q ) \rangle  =
\mid R^{0}_{\alpha \beta} (\PP ,\q ) \rangle +
\frac{1}{\E{q}\pm i\delta -\hat{H}}
\;\hat{H}_I\;\mid R^{0}_{\alpha \beta} (\PP ,\q ) \rangle.
\label{Reslvsol}
\ee
It is a simple matter to see that all integral equations above are exactly
solvable for degenerate kernels:
$
K^{\PP}_{22}(\rv ,\s )= \sum_n V_n(\rv )\;U_n(\s ).
$

The dynamical mapping formulae (\ref{dm1}),(\ref{dm2}) give two forms of
instantaneous Bethe-Salpeter matrix element
$
\langle 0\mid a_{\alpha}(\frac{\PP}{2}+\kap,t)\,
a_{\beta}(\frac{\PP}{2}-\kap,t)\,\mid 2,\PP \rangle,
$
leading to the following identities when $\mid 2,\PP \rangle$ stands for
$\mid R^{\pm}_{\alpha \beta} (\PP,\q ) \rangle $ or
$\mid {\rm B}^{\PP}_{\alpha \beta}\rangle $ :
\bea
\left. \begin{array}{c}
\VF{\kap}\left[ e^{it\left(\E{\kappa}-\E{q}\mp i\delta\right)}-1\right] \\
              \\ \displaystyle
B^{\PP}(\kap)\left[ e^{it\left(\E{\kappa}-M_2(\PP)\right)}-1\right]
\end{array}\right\}=
\di d^3r \left\{ \begin{array}{c}
\VF{\rv} \\   \\ \displaystyle  B^{\PP}(\rv) \end{array} \right\}\cdot &&
\nonumber \\
\cdot
F^{(1)}_A\left(t;\kap+\rv \mid\frac{\PP}{2}-\kap ,\frac{\PP}{2}-\rv \right),
&&
\label{OffShlUn}
\eea
which have a sense of off-shell extension of unitarity relation .

\begin{center}
{\bf 2. Four-fermion models }
\end{center}

As a first example let us consider the contact four-fermion model.
Defining
\bea
&& x=(\x,t);\;\; t=x_0;\;\;\;{\Pop}_{\rm x}=-i\vec{\nabla}_{\rm x};
\;\;\epsilon_{\alpha\beta}=-\epsilon_{\beta\alpha};\;\;\;
\chi_{(\Psi)}(x)=\epsilon_{\alpha\beta}\,\Psi_{\alpha }(x)\,
\Psi_{\beta}(x);
\nonumber \\
&& S_{(\Psi)}(x)=\Psi^{\dagger}_{\alpha }(x)\,\Psi_{\alpha }(x),\;\;\;\;\;
{\jv}_{(\Psi)}(x)=\Psi^{\dagger}_{\alpha }(x)
\stackrel{\longleftrightarrow}{{\Pop}}\Psi_{\alpha }(x);
\label{CONV} \\
&& \left\{\Psi_{\alpha }(x)\,,\Psi_{\beta}(y)\right\}\Biggr|_{x_0=y_0}=0;
\;\;\;\mbox{\bf with the convention:}
\nonumber \\
&& \left\{\Psi_{\alpha }(x)\,,\Psi^{\dagger}_{\beta}(y)\right\}
\Biggr|_{x_0=y_0}=\delta_{\alpha\beta}\,\delta_3(\x-\y)\Rightarrow
\Biggr|_{\x=\y} \delta_{\alpha\beta}\;\frac{1}{V^*},
\nonumber
\eea
let us consider the local 4-fermion interactions with the following
densities:
\bea
&& {\cal H}_1(x)=\Psi^{\dagger}_{\alpha }(x)\,E({\Pop})\,\Psi_{\alpha }(x) -
\frac{\lambda}{8}\  \chi^{\dagger}_{(\Psi)}(x)\ \chi_{(\Psi)}(x);
\label{H1} \\
&& {\cal H}_2(x)=\Psi^{\dagger}_{\alpha }(x)\,{\cal E}({\Pop})\,
\Psi_{\alpha }(x) -\frac{\lambda}{4}\  S^2_{(\Psi)}(x)+\frac{\mu}{4}\
{\jv}^2_{(\Psi)}(x) \Rightarrow
\label{H2} \\
&& :{\cal H}_2(x):=w_0+\Psi^{\dagger}_{\alpha }(x)\,E({\Pop})\,
\Psi_{\alpha }(x) - \frac{\lambda}{8}\  \chi^{\dagger}_{(\Psi)}(x)\
\chi_{(\Psi)}(x)+
\nonumber \\
&& + \frac{\mu}{4}\,\Psi^{\dagger}_{\alpha }(x)\,
\left({\jv}_{(\Psi)}(y)\cdot
\stackrel{\longleftrightarrow}{\Pop}_{\rm x}\right)\Psi_{\alpha }(x)
\Biggr|_{x=y},
\nonumber
\eea
which give the HE (\ref{Geq?}),(\ref{Geq??}) for HF:
\bea
&& \Psi_{\alpha}(x) =\di\frac{d^3k}{(2\pi)^{3/2}}\, e^{i(\ka\cdot\x)-iE(k)t}
\, a_{\alpha}(\ka,t),
\label{Psi} \\
&& \left(i\partial_t-E({\Pop}) \right)\ \Psi_{\alpha}(x) =
\hat{\rm V}_{(\Psi)}(t;\x,\Pop)\, \Psi_{\alpha}(x) \equiv
{\cal J}_{\alpha}(x),
\label{PsiEQ} \\
&& \hat{\rm V}_{(\Psi)}(t;\x,\Pop)=-\frac{\lambda}{2}S_{(\Psi)}(x)+
\mu\left[\left({\jv}_{(\Psi)}(x)\cdot\Pop_{\rm x}\right)-\frac{i}{2}
\left(\vec{\nabla}_{\rm x}\cdot {\jv}_{(\Psi)}(x)\right) \right];
\nonumber \\
&& -i<\ka|e^{itE({\Pop})}
\hat{\rm V}_{(\Psi)}(t;\Xop,\Pop)e^{-itE({\Pop})}|\lv>=
-i\,e^{it[E(k)-E(l)]}\,<\ka|\hat{\rm V}_{(\Psi)}(t;\Xop,\Pop)|\lv>
\nonumber \\
&& = Q_{(a)}(\ka,\lv;t)=\di d^3\PP \,a^{\dagger}_{\beta}(\PP -\ka,t)\,
a_{\beta}(\PP -\lv,t)\,e^{it[\E{\frac{\PP}{2}-\ka}-\E{\frac{\PP}{2}-\lv}]}
\cdot
\nonumber \\
&& \cdot\frac{2}{i}K^{\PP}_{22}\left(\lv-\frac{\PP}{2},\ka-\frac{\PP}{2}
\right),
\nonumber
\eea
and for the following simple operator realizations via physical fields
\bea
&& \Psi_{\alpha}(\x,0)=\Psi_{\alpha}\left[\psi_{\alpha}\left\{ A(\ka)\right\}
\right] \Rightarrow \left\{
\begin{array}{c}
 -\epsilon_{\alpha\beta}\,\psi^{\dagger}_{\beta}(\x,0);\;\;\;(+), \;\;
E(k)\Rightarrow E^+(k) \\
\psi_{\alpha}(\x,0);\quad (-),\;\; E(k)\Rightarrow E^-(k)
\end{array} \right.
\nonumber \\
&& \psi_{\alpha}(x) = \di \frac{d^3k}{(2\pi)^{3/2}} \, e^{i(\ka\cdot\x)-iE(k)t}
A_{\alpha}(\ka)=e^{-itE(\Pop)}\psi_{\alpha}(\x,0),
\label{psi}
\eea
lead to above reduced Hamiltonian (\ref{Hred})
$H\{A\}=:H_2\{A\}:$
(note, that ${\cal H}_1(x)$ looks like normal form of ${\cal H}_2(x)$
for the case $\mu=0$), with the following degenerate kernel and parameters:
\bea
&& \frac{2}{i} K^{\PP}_{22} (\rv ,\s ) \Rightarrow
\frac{i}{2 (2\pi )^3 } \left \{L + \mu\left(\rv +\s \right)^2 \right \};
\quad  L\equiv \lambda -\mu {\PP}^2;
\label{K22} \\
&& E(k)\Rightarrow E^{\pm}(k)=\frac{\mu}{4V^*}\left(k^2+<k^2>\right)+
\frac{\lambda}{4V^*}\mp \left({\cal E}(k)-\frac{\lambda}{2V^*}\right);
\label{E1pm} \\
&& w_0\Rightarrow w^{\pm}_0=\frac{1 \pm 1}{V^*}
\left(<{\cal E}(k)>-\frac{\lambda}{2V^*}\right);\;\;
<{\cal E}(k)>\stackrel{def}{=}V^*\di \frac{d^3k}{(2\pi)^3} \,{\cal E}(k).
\nonumber
\eea
Here ${\cal E}(k)$ is arbitrary "bare" one-particle spectrum and
$V^*$ has a sense of the volume of excitation.

Now the solution of (\ref{Feq}) for the first coefficient function reads :
\bea
&& F^{(1)}\left(t;\kap+\q \mid \frac{\PP}{2}-\kap ,\frac{\PP}{2}-\q \right)=
\di \limits^{t}_0 d\eta \,e^{i\eta\E{q} }
\frac{\varepsilon (t)}{2\pi i}\di
\limits^{i\infty+\Delta\cdot\varepsilon(t)}_{-i\infty+
\Delta\cdot\varepsilon(t)}
\frac{d\sigma \,e^{\sigma \eta }}{\sigma+i\E{\kappa}}\cdot
\nonumber \\
&& \cdot \left\{\frac{\D{\kappa\,1}+ q^2\,\D{\kappa\,2}}{\D{0}}+
2\mu\,q^j \left[\frac{{\prod}^{jl}_{\perp}(\PP)}{ 1-2\mu \J{\delta} }+
\frac{{\prod}^{jl}_{\parallel}(\PP)}{ 1-2\mu \J{D} } \right] \kappa^l\right\},
\label{FSol}
\eea
where $\varepsilon (\eta)=\varepsilon (t) \equiv $ signum$(t)$, $\Delta >0$,
and
\bea
&& \left\{\begin{array}{c}
\J{n} \\ \J{D} \\ \J{jl} \end{array}\right\}=
\frac{1}{2} \di \frac{d^3r}{(2\pi)^3}\,\cdot \frac{1}{\E{r}-i\sigma}\cdot
\left\{\begin{array}{c} \displaystyle\left(r^2\right)^n   \\
\displaystyle \frac{\left(\rv\cdot\PP\right)^2}{\PP^2}  \\
\displaystyle  r^j r^l \end{array}\right.  ;
\label{JJJ} \\
&& \J{jl}=\delta_{jl}\J{\delta}+\frac{\PP^j\PP^l}{\PP^2}\J{R},\quad
\J{\delta}=\frac{1}{2}\left[\J{1}-\J{D}\right];
\nonumber \\
&& \left.\begin{array}{c}\D{0} \\  \D{\kappa \,1} \\
 \D{\kappa \,2}  \end{array}\right\}= \left\{
\begin{array}{c}  \displaystyle
\left[\mu\,\J{1}-1\right]^2- \mu^2\,\J{0}\J{2}-L\,\J{0}.\\
\displaystyle
L+\mu \kappa^2+\mu^2\left[\J{2}-\kappa^2\J{1}\right]. \\
\displaystyle
\mu +\mu^2 \left[\kappa^2\J{0}-\J{1}\right].
\end{array} \right. .
\label{DDD}
\eea
Setting $\I{\{\ldots\}}\equiv {\rm J}^{\cal P}_{\{\ldots\}}(\pm\delta-i\E{q}),
\quad \DD{\{\ldots\}}=D^{\cal P}_{\{\ldots\}}(\pm\delta-i\E{q})$,
from eq.(\ref{LpmnSchwgr}) or from (\ref{Scatt_F}) for the scattering
eigenfunctions with fixed spin J=0,1, defined in symmetrical basis
($\sigma_j,\; j=1,2,3$ -usual Pauli matrices):
\[
(\delta_{\alpha \beta},(\sigma_j)_{\alpha \beta})\longrightarrow
(\epsilon_{\alpha \beta},\tau^j_{\alpha \beta}), \;\;\;
\epsilon_{\alpha \beta}=i(\sigma_2)_{\alpha \beta},\;\;\;
\tau^j_{\alpha \beta}=\tau^j_{\beta\alpha}=
i(\sigma_2\sigma_j)_{\alpha \beta},
\]
\bea
&& \mbox{as:}\qquad\;\;\vf{0}{\kap}=\epsilon_{\alpha \beta}\;
\Phi^{\pm(0)}_{{\cal P}q}(\kap);\qquad
\vf{1,m}{\kap}=\tau^{(m)}_{\alpha \beta}\;\Phi^{\pm(1)}_{{\cal P}q}(\kap);
\label{SpSt} \\
&& \tau^{(0)}_{\alpha \beta}=i\tau^{3}_{\alpha \beta};\quad
\tau^{(\pm 1)}_{\alpha \beta}=\frac{\mp i}{\sqrt{2}}\left(
\tau^{1}_{\alpha \beta}\pm i\tau^{2}_{\alpha \beta}\right);
\qquad \vf{\rm J,m}{\kap}=-\phi^{\pm ({\rm J,m})}_{{\cal P}q}
(-\kap)_{\beta\alpha};
\nonumber \\
&& \mid J,m;\PP,q\rangle =
\di d^3\kappa\,\vf{\rm J,m}{\kap}\,\mid R^0_{\alpha\beta}(\PP,\kap )\rangle;
\nonumber
\eea
one has the following expressions:
\bea
&& \Phi^{\pm(\rm J)}_{{\cal P}q}(\kap)=\frac{1}{2}\left[\VF{\kap}+
(-1)^{\rm J} \VF{-\kap}\right];
\label{PhiJ} \\
&& \Phi^{\pm(\rm J)}_{{\cal P}q}(\kap)=\frac{1}{2}\left[\delta_3(\kap-\q)+
(-1)^{\rm J} \delta_3(\kap+\q)+
\frac{\TT{\rm J}{\kappa} }{\E{\kappa}-\E{q}\mp i0 }\right];
\nonumber \\
&& \TT{0}{k}=\CC{1}+k^2\,\CC{2}=\frac{ \DD{q1}+k^2\,\DD{q2} }
{(2\pi)^3 \DD{0} },
\label{TT0} \\
&& \TT{1}{k}= \left(\ka \cdot \CC{3}\right) =
\frac{2\mu }{(2\pi)^3} k^j\left\{
\frac{{\prod}^{jl}_{\perp}(\PP)}{ 1-2\mu \I{\delta} }+
\frac{{\prod}^{jl}_{\parallel}(\PP)}{ 1-2\mu \I{D} } \right\} q^l;
\nonumber \\
&& \mbox{where projectors are}\;\;
{\prod}^{jl}_{\perp}(\PP)=\left(\delta_{jl}-\frac{\PP^j\PP^l}{\PP^2}
\right); \quad {\prod}^{jl}_{\parallel}(\PP)=\frac{\PP^j\PP^l}{\PP^2}.
\label{TT1}
\eea
The bound state wave functions look like simple residues of scattering one
$ \Phi^{\pm(\rm J)}_{{\cal P}q}(\ka) $ for corresponding poles for
$\E{q}\pm i\delta\Rightarrow i\sigma \Rightarrow M^{(\rm J)}_2(\PP);$
\bea
&& \D{0}=0; \quad 1-2\mu \J{\delta}=0; \quad 1-2\mu \J{D}=0;
\;\;\;C_n(\PP)\sim \CC{n};
\nonumber \\
&&\Phi^{\pm(\rm J)}_{{\cal P}q}(\ka)\Rightarrow B^{\PP ({\rm J})}(\ka)=
Z^{(\rm J)}(\PP,\ka) \left[\E{k}-M^{(\rm J)}_2(\PP)\right]^{-1};
\label{bndst} \\
&& Z^{(0)}(\PP,\ka)=C_1(\PP)+k^2\,C_2(\PP);\quad
Z^{(1)}_{\perp,\parallel}(\PP,\ka)=\left(\ka \cdot
\vec{\rm C}^{\perp,\parallel}_3(\PP)\right);
\nonumber \\
&&  \mbox{For the case} \quad\mu\rightarrow 0:\quad
\CC{1}\Rightarrow \frac{\lambda (2\pi)^{-3}}{\lambda\I{0}-1};\qquad
\CC{2,3}\Rightarrow 0.
\label{mu0}
\eea
The obtained solutions directly satisfy to extended unitarity
relation (\ref{OffShlUn}).

\begin{center}
{\bf 3. Case $\mu=0$ and linearisation of HE}
\end{center}

Returning to HE, let us consider the conserved charge
densities $S_{(\Psi)}(x)$:
\bea
&& i\partial_t S_{(\Psi)}(x)=\Psi^{\dagger}_{\alpha }(x)\left[
E(\stackrel{\rightarrow}{\Pop})-E(\stackrel{\leftarrow}{\Pop})\right]
\Psi_{\alpha }(x)
-i\mu \,\vec{\nabla}_{\rm x}\cdot \left(\Psi^{\dagger}_{\alpha }(x)\,
 {\jv}_{(\Psi)}(x) \Psi_{\alpha }(x)\right);
\nonumber \\
&& Q(t)=\di d^3{\rm x}\,S_{(\Psi)}(\x,t);\;\;\;
\partial_t\; Q(t)=-\mu \oint \limits_{\Sigma_R} d\vec{\sigma}\cdot
\left(\Psi^{\dagger}_{\alpha }(x)\, {\jv}_{(\Psi)}(x) \Psi_{\alpha }(x)\right)
\stackrel{R\rightarrow\infty}{\rightarrow}0.
\nonumber
\eea
It is clear that for $\mu =0$ the HE for $S_{(\Psi)}(x)$ contains only
kinetic term. Then the initial conditions lead to the simple form of
dynamical mapping for this operator:
\bea
&& S_{(\Psi)}(\x,t)\equiv e^{i Ht}\, S_{(\Psi)}(\x,0) \,e^{-i Ht}
\Rightarrow e^{iH_0t}\, S_{(\Psi)}(\x,0) \,e^{-iH_0t}
\Rightarrow
\nonumber \\
&& \Rightarrow e^{iH_0 t}\, S_{(\psi)}(\x,0) \,e^{-iH_0 t}
\equiv S_{(\psi)}(\x,t).
\label{Sdm}
\eea
So, HE (\ref{PsiEQ}) become {\sl linear} equation with respect to HF
$\Psi_{\alpha}(\x,t) $:
\be
\left(i\partial_t-E({\Pop}) \right)\ \Psi_{\alpha}(\x,t) =
-\frac{\lambda}{2}S_{(\Psi)}(\x,t)\,\Psi_{\alpha}(\x,t)\Rightarrow
-\frac{\lambda}{2}S_{(\psi)}(\x,t)\,\Psi_{\alpha}(\x,t),
\label{LinHE}
\ee
and its solution gives the closed expression of HF in terms of the
physical one:
\bea
&& \Psi_{\alpha}(\x,t)=\mbox{\bf T}\;\exp\left\{i \di \limits^t_0 d\eta \left[
\frac{\lambda}{2} S_{(\psi)}(\x,\eta)-E(\Pop)\right]\right\}
\psi_{\alpha}(\x,0)=
\nonumber \\
&& =e^{-itE(\Pop)}\mbox{\bf T}\;\exp\left\{i\,\frac{\lambda}{2}\di
\limits^t_0 d\eta \, S_{(\psi)}\left(\x+\eta \vs(\Pop),\eta \right)\right\}
\psi_{\alpha}(\x,0),
\label{PsiSoltn}
\eea
where $\vs(\Pop)=\vec{\nabla}_{\mbox{\bf p}}E(\Pop)$ is corresponding
group velocity. Dynamical mapping is given by normal ordering of
this formal solution. It seems difficult to obtain such kind of solution
in terms of in-fields.

\begin{center}
{\bf 4.$N,\Theta$ model}
\end{center}

This model is determined by the following Hamiltonian, CCR, and HE:
\bea
&& H=\di d^3{\rm x} \Biggl \{N^{\dagger} (x) \mu (\nabla^2) N(x) +
\Theta^{\dagger} (x) w (\nabla^2)\Theta (x) +
\nonumber  \\
&& + \lambda \di d^4y \di d^4z \bar{\alpha
} (x-y) \bar{\alpha } (x-z)
N^{\dagger}(x) N(x) \Theta^{\dagger} (y) \Theta (z) \Biggr \},
\label{ham} \\
&& \bar{\alpha } (x-y) =\alpha ({\rm x}-{\rm y}) \delta (t_x-t_y);\;\;
\mu (\nabla^2) e^{ik {\rm x}} = m(k) e^{ik {\rm x}};\;\;
w (\nabla^2) e^{ik {\rm x}}  = \omega (k) e^{ik {\rm x}}.
\nonumber \\
&& \{N(x), N^{\dagger }(y)\} \delta (t_x-t_y) = \delta_4 (x-y),\quad
[\Theta (x), \Theta^{\dagger } (y)] \delta (t_x-t_y) = \delta_4 (x-y),
\nonumber \\
&& (i\partial_t-\mu (\nabla^2)) N(x)=
\lambda \di d^4y \di d^4z \bar{\alpha }
(x-y) \bar{\alpha } (x-z) \Theta^{\dagger} (y) N(x) \Theta (z),
\nonumber \\
&& (i\partial_t-w (\nabla^2)) \Theta (x)=
\lambda \di d^4y \di d^4z \bar{\alpha }
(y-x) \bar{\alpha } (y-z) N^{\dagger} (y) N(y) \Theta (z).
\label{heq}
\eea
All others (anti) commutators vanish. It is seen from this HE that as
in the previous case (\ref{Sdm}) the HE for the operator $N^{\dagger}(x)N(x)$
contains only kinetic term:
\be
i\partial_t(N^{\dagger}(x)N(x))=N^{\dagger}(x)\left[
\mu(\stackrel{\rightarrow}{\nabla^2}) -
\mu(\stackrel{\leftarrow}{\nabla^2}) \right]N(x).
\label{NNeq}
\ee
Then $ N^{\dagger}(x)N(x) = N^{\dagger}_0(x)N_0(x)$. Defining the HF
$N(x)$ and physical field $N_0(x)$ and HE in momentum representation
\bea
&& \left.
\begin{array}{c}
N(x) \\ \Theta (x)
\end{array}\right\}
 = \di \frac{d^3k}{(2\pi)^{3/2}}\left\{
\begin{array}{c}
n(k, t)\,e^{ik {\rm x}-im(k)t} \\ o(k, t)\,e^{ik {\rm x}-i\omega(k)t}
\end{array}\right\} ;\;\;\;
\tilde{\alpha }(p) = \di \frac{d^3{\rm x}}{(2\pi)^{3/2}} e^{ip {\rm x}}
\alpha ({\rm x});
\label{hflds} \\
&& i\partial_t n(l,t)=\lambda \di d^3p \di d^3q\, \tilde{\alpha } (q)
\tilde{\alpha } (-p) e^{i(E^{q+l}_q - E^{q+l}_p)t}
o^{\dagger}(q,t)o(p,t)n(l+q-p,t),
\nonumber \\
&& i\partial_t o(l,t)=
\label{heqtn} \\
&& = \lambda \tilde{\alpha }(l)  \di d^3p \di d^3q \,\tilde{\alpha }(p-q-l)
e^{i(E^{q+l}_l -E^{q+l}_{l+q-p})t} n^{\dagger}(q,t)n(p,t)o(l+q-p,t),
\nonumber
\eea
where $E^{p+q}_q =\omega (q)+m(p)$; and $N_0(x),\Theta_0(x)$ are defined by
the same identities (\ref{hflds}) with $n(k,t)\rightarrow N(k)$,
$ o(k,t)\rightarrow \Theta(k) $, one has, as above, the linear HE
for operator $o(k,t)$
\bea
&& o(k,t)=\Theta (k) + \di \limits^t_0 d\eta \di d^3l \;{\cal K}_N(k,l;\eta)
o(l,\eta),
\label{INTheq} \\
&& {\cal K}_N(k,l;t)=-i\lambda \tilde{\alpha }(k)\tilde{\alpha }(-l)\di d^3q
\,e^{i(E^{q+k}_k -E^{q+k}_{l})t} N^{\dagger}(q) N(q+k-l),
\nonumber
\eea
with initial conditions $o(k,0)=\Theta (k),\;\;n(k,0)=N(k)$
which has the similar formal solution
\be
o(k,t)=\Theta (k)+ \di d^3l \; {\cal R}_N(k,l;t) \Theta (l)=
{\rm T} \left\{\exp\left[\di \limits^t_0 d\eta {\hat K}_N(\eta)\right]
\Theta \right\}(k),
\label{frmsltn}
\ee
where ${\hat K}_N(\eta)$ is integral operator with the kernel
${\cal K}_N(k,l;\eta)$.
Note, that for a given $o(k,t)$ the equation (\ref{heqtn}) for $n(k,t)$
is also linear.

For the reduced Hamiltonian
\bea
&& H_I=\lambda \di d^3p \di d^3k \di d^3q \tilde{\alpha }(-q)
\tilde{\alpha }(k+q-p)\,N^{\dagger}(p)N(k)\Theta^{\dagger} (k+q-p) \Theta (q),
\nonumber \\
&& H_0=\di d^3p \,\left[m(p)\, N^{\dagger}(p)N(p) + \omega (p)\,
\Theta^{\dagger}(p) \Theta (p)\right],\quad
H=H_0 + H_I,
\label{redham}
\eea
as for the previous case, it's possible to find coefficient functions of
dynamical mapping and two-particles bound and scattering eigenstates.
The dynamical mapping up to third order now reads :
\bea
&& o(p,t)={\cal O}^t_0[\Theta(k),N(k)]=
\Theta (p)+\di d^3q \di d^3k N^{\dagger}(q) N(p+q-k) \Theta (k)\cdot
\nonumber  \\
&& \cdot F(t;q,p;p+q-k,k)+\ldots
\nonumber  \\
&& n(p,t)={\cal N}^t_0[\Theta(k),N(k)]=
N(p)+\di d^3q \di d^3k \Theta^{\dagger}(q) N(p+q-k) \Theta (k)\cdot
\nonumber  \\
&& \cdot F(t;p,q;p+q-k,k)+\ldots ,
\label{DMon}
\eea
where the first coefficient function may be found as:
\bea
&& F(t;l,q;l+q-p,p) = -i\lambda \tilde{\alpha }(q)\tilde{\alpha }(-p) \di
\limits^t_0 d\xi e^{i\xi E^{l+q}_q}
\frac{\varepsilon (t)}{2\pi i}\cdot
\label{FFF1} \\
&& \cdot\di
\limits^{i\infty+\Delta\cdot\varepsilon (t)}_{-i\infty+
\Delta\cdot\varepsilon (t)}
d\sigma \frac{e^{\sigma \xi}}{(\sigma + iE^{l+q}_p)
\left [ 1+ I^{l+q}(\sigma )\right ] };
\quad I^{l+q}(\sigma )=
\lambda \di d^3 k \frac {\mid \tilde{\alpha } (k)\mid^2 }{
E^{l+q}_k - i\sigma }.
\nonumber
\eea
The familiar solutions of eq. (\ref{LpmnSchwgr}) for bound and scattering
eigenstates of Hamiltonian (\ref{redham}) \cite{Umz},\cite{Schweb} :
\bea
&& H\mid R^{\pm}\{N(p-q) \Theta (q)\} \rangle =
 E^{p}_q \mid R^{\pm}\{N(p-q) \Theta (q)\} \rangle ,\;\;
H \mid {\rm B}^{\PP} \rangle =M(\PP)\mid {\rm B}^{\PP} \rangle;
\nonumber  \\
&& \mid R^{\pm}\{ N(p-k)\Theta (k)\} \rangle = \di d^3q R^{p,k}_{\pm}(q)\,
N^{\dagger}(p-q)\Theta^{\dagger} (q)\mid 0\rangle,
\label{r} \\
&& R^{p,k}_{\pm}(q) = \delta_3 (k-q)+ \frac{\lambda \tilde{\alpha }(q)
\tilde{\alpha }(-k)}{E^p_{k}-E^p_{q} \pm i\delta} Q^{p(\pm)}_k;\;\;
Q^{p(\pm)}_k=\left[1+ I^{p}\left(\pm \delta-iE^p_{k})\right)
\right]^{-1};
\nonumber  \\
&& \mid {\rm B}^{p} \rangle =\di d^3q B^{p}(q)\;
N^{\dagger}(p-q)\Theta^{\dagger}(q)\mid 0\rangle; \;\;\;
1+I^{p}\left(-iM(p)\right)=0.
\label{bond} \\
&& B^p (q) = Z^p \frac {\tilde{\alpha }(q)}{M (p) - E^p_{q}};\;\;
J^{p}(\sigma )=\di d^3 k \frac {\mid \tilde{\alpha } (k)\mid^2 }
{(E^{p}_k - i\sigma)(E^{p}_k - M(p)) };\;\;
\nonumber  \\
&& \di d^3k \stackrel{*}{B^p}(k) B^p(k)= \mid Z^p \mid^2 J^{p}(-iM(p))=1,
\label{Norm1}
\eea
satisfy to the orthogonality conditions:
\bea
&& \langle R^{\pm}\{ N(p-k)\Theta (k)\} \mid R^{\pm}\{ N(p_1-k_1)
\Theta (k_1)\} \rangle =
\nonumber \\
&& =\delta_3(p-p_1)\di d^3q
\stackrel{*}{R^{p,k}_{\pm}}(q)\, R^{p,k_1}_{\pm}(q) =
\delta_3(p-p_1)\, \delta_3(k-k_1);
\label{nrmSCT} \\
&& \langle {\rm B}^{p_1} \mid R^{\pm}\{N(p-k) \Theta (k)\} \rangle =
\delta_3(p_1-p) \di d^3 q B^{p*}(q) R^{p,k}_{\pm}(q)=0.
\nonumber
\eea
By definition, the S-matrix reads:
\bea
&& \langle N_{in}(p-k)\Theta_{in}(k) \mid \hat{\rm S}^{\pm 1}
\mid N_{in}(p_1-k_1)\Theta_{in}(k_1) \rangle \stackrel{def}{=}
\nonumber  \\
&& \stackrel{def}{=}
\langle R^{\pm}\{ N(p-k)\Theta (k)\} \mid R^{\mp}\{ N(p_1-k_1)
\Theta (k_1)\} \rangle =
\nonumber  \\
&& =\delta_3(p-p_1)\di d^3q
\stackrel{*}{R^{p,k}_{\pm}}(q)\, R^{p,k_1}_{\mp}(q) =
\delta_3(p-p_1)\,S^p_{\pm}(k,k_1);
\label{Smtr} \\
&& S^p_{\pm}(k,k_1)= \left\{\delta_3(k-k_1)\mp 2\pi i\,
\delta(E^p_{k_1}-E^p_k)\,
\lambda \,\tilde{\alpha }(-k_1) \tilde{\alpha}(k)\,Q^{p(\pm)}_k \right\}.
\nonumber
\eea
This model was considered also by Umezawa, Matsumoto, Tachiki
\cite{Umz} in the framework of dynamical mapping method using the
"in" physical fields. The dynamical mapping for this case looks like this:
\bea
&& n(l,t) \equiv {\cal N}^t_{-\infty}[N_{in}(k),\Theta_{in}(k)]=N_{in}(l)+
\nonumber \\
&& +\di d^3q\di d^3k\left[R^{l+q,k}_{\pm}(q)-\delta_3(k-q)\right]\,
e^{i(E^{l+q}_q-E^{l+q}_k)t}\Theta^{\dagger}_{in}(q) N_{in}(l+q-k)
\Theta_{in}(k)+
\nonumber \\
&& +\di d^3k B^{l+k}(k)\,e^{i(E^{l+k}_k - M(l+k))t}
\Theta^{\dagger}_{in}(k) V_{in}(l+k)+...\; ;
\nonumber \\
&& e^{-itE^p_q } o(q,t) n(p-q,t) =
e^{-itM(p)}\,B^p(q)\,V_{in}(p)+
\nonumber \\
&&+\di d^3k\,e^{-itE^p_k }\,R^{p,k}_{+}(q)\,
\Theta_{in}(k)N_{in}(p-k)+\ldots .
\label{onDM}
\eea
Unfortunately, the second term in the last equation was omitted in \cite{Umz}.
With this correction, one can compare their results with the our approach
using the uniqueness of HF and making the dynamical mapping onto the
"in" field by two steps:
\be
n(k,t)={\cal N}^t_{-\infty}[N_{in}(k)]={\cal N}^t_0[N(k)]=
{\cal N}^t_0\left\{{\cal N}^0_{-\infty}[N_{in}(k)]\right\}.
\label{Comp}
\ee
One can see that obtained consistency conditions have the same form as
above mentioned off-shell extended unitarity relations (\ref{OffShlUn}):
\bea
&& B^{l+q}(q) \left \{ e^{i(E^{l+q}_q - M(l+q))t} - 1 \right \} =
\di d^3p\,B^{l+q} (p) F(t;l,q;l+q-p,p),
\label{BR} \\
&& R^{l+q,k}_{\pm}(q) \left \{ e^{i(E^{l+q}_q - E^{l+q}_k)t} - 1 \right \} =
\di d^3p\, R^{l+q,k}_{\pm}(p) F(t;l,q;l+q-p,p),
\nonumber
\eea
and hold identically for solutions (\ref{FFF1}-\ref{Norm1}).
Moreover, the scattering and bound states for this two different
approaches are connected correspondingly as:
\bea
&& \langle{\rm B}^p \mid =\langle 0 \mid \di d^3q\,
B^{p*}(q)\,N(p-q) \Theta (q)=
\nonumber \\
&& = \langle 0 \mid \di d^3q \, B^{p*}(q)\,
\left\{B^p(q)\,V_{in}(p)+\di d^3k\,R^{p,k}_{+}(q)\,\T(k)\N(p-k)
+\ldots\right\}=
\nonumber  \\
&& =\langle 0 \mid V_{in}(p);
\nonumber  \\
&& \mid R^{\pm}\{ N(p-k)\Theta (k)\} \rangle = \di d^3q R^{p,k}_{\pm}(q)\,
N^{\dagger}(p-q)\,\Theta^{\dagger }(q)\mid 0\rangle =
\nonumber \\
&& =\di d^3q R^{p,k}_{\pm}(q)\,\left\{\stackrel{*}{B^p}(q)\,
V^{\dagger}_{in}(p)+
\di d^3l\,\stackrel{*}{R^{p,l}_{+}}(q)\,\NM(p-l)\TY(l)+\ldots\right\}
\mid 0\rangle =
\nonumber  \\
&& =\di d^3l\,\NM(p-l)\TY(l)\,\mid 0 \rangle \left\{\begin{array}{c}
\delta_3(l-k),\quad (+)   \\ S^p_+(l,k) \quad (-).
\end{array}\right.
\label{vo}
\eea
So, as expected, the bound state $\mid {\rm B}^p \rangle $, obtained by
selfconsistency method of \cite{Umz} with the help of new bound state
operator $V_{in}(p)$ coinsides with the one obtained from direct solution
of eigenvalue problem in terms of constituent fields, and in their turn
the scattering eigenstates are nothing but two-particles in- and
out- states from \cite{Umz}.

\begin{center}
{\bf 5. Divergencies and selfadjoint extensions }
\end{center}

Now some remarks about divergency problem are needed.
It's absent for $N,\Theta$ model due to $\tilde{\alpha}(p)$. However for
four-fermion models with quadratic "bare" spectra ${\cal E}(k)$ the
two-particle eigenvalue problem (\ref{LpmnSchwgr}) may be reduced in
configuration space to the Schroedinger equation with singular delta-like
potential, considered in \cite{BrFddShr}. A simple cut-off procedure for
integrals (\ref{JJJ}) with forthcoming incorporation of cut-off parameter
and "bare" coupling constant into the binding energy $M_2(0)$ via
subtraction procedure leads to the same answer where arbitrary binding
energy serves as a parameter of selfadjoint extension of free Hamiltonian
$(-\nabla^2_{\rm x})$ in two-particle sector \cite{BrFddShr}.

The example of such dimensional transmutation of cut-off parameter and "bare"
coupling constant into the binding energy for 2D Schroedinger
equation was given in \cite{Thorn}. So doing, we obtain from (\ref{PhiJ}),
(\ref{TT0}), (\ref{mu0}), for Hamiltonian ${\cal H}_1$ (\ref{H1}) with
${\cal E}(k)=k^2/2m+{\cal E}_0$, $\lambda_0=\lambda m/2$, and
$M_2(\PP)=\PP^2/4m+2{\cal E}_0-b^2/m$
the following renormalized eigenfunctions:
\be
\VF{\ka}=\delta_3(\ka-\q) -
\frac{\left[2\pi^2\,(b\pm iq)\right]^{-1}}{(k^2-q^2\mp i0)};
\qquad B(k)=\frac{\sqrt{8\pi b}}{k^2+b^2};
\label{FB1}
\ee
where the bound state equation (\ref{bndst}) leads to transmutation
condition:
\[
1=\lambda\; {\rm J}^{\PP}_0\left(-iM_2(\PP)\right)\Rightarrow\lambda_0
\di \limits_{k<\Lambda}\frac{d^3k}{(2\pi)^3}\cdot\frac{1}{k^2+b^2}\Rightarrow
\frac{\lambda_0}{4\pi}\left(\frac{2}{\pi}\Lambda-b\right);
\quad b=\frac{2}{\pi}\Lambda-\frac{4\pi}{\lambda_0}.
\]
The Fourier-images $\phi(\x)$ of wave functions (\ref{FB1}) satisfy to
Schroedinger equation with singular delta-potential:
\[
\left(-\nabla^2 +b^2\right)\phi_B(\x)=\lambda_0\,\delta_3(\x)\,\phi_B(\x).
\]
So, the existence of selfadjoint extension of free Hamiltonian corresponding
to such singular one implies definite behavior of "bare" coupling constant
as well as in \cite{BrFddShr}:
\be
\lambda_0(\Lambda)=4\pi\left[\frac{2}{\pi}\Lambda-b\right]^{-1}
\simeq\frac{2\pi^2}{\Lambda}\left(1+\frac{\pi b}{2\Lambda}+\ldots
\right).
\label{lmd0}
\ee
The case of Hamiltonian (\ref{H2}) becomes less trivial because it requires
renormalization of mass and "gap" of one-particle spectrum (\ref{E1pm})
as well:
\bea
&& E^{\pm}(k)=\frac{k^2}{2{\cal M}^{\pm}}+E^{\pm}_0;\quad
E^{\pm}_0=g\left(\frac{<k^2>}{(2mc)^2}+1\right)\pm (2g-{\cal E}_0);
\nonumber  \\
&& \mbox{where we put:}\;\;\mu\Rightarrow\frac{\lambda}{(2mc)^2};\quad
g=\frac{\lambda}{4V^*};\quad {\cal M}^{\pm}=
m\left(\frac{g}{2mc^2}\mp 1\right)^{-1}.
\label{FB2}
\eea
The dependence of "bare" quantities on cut-off parameter $\Lambda$ is now
fixed as:
\bea
&& \frac{1}{V^*}=\frac{\Lambda^3}{6\pi^2};\;\;<k^2>=\frac{3}{5}\Lambda^2;
\;\;g=\Lambda^2G(\Lambda);\;\;(2mc)^2=\Lambda^2\nu(\Lambda);\;\;
{\cal E}_0=\Lambda^2\epsilon(\Lambda);
\nonumber  \\
&& 2{\cal M}^{\pm}\equiv 2{\cal M}^{\pm}(\Lambda)=
\frac{\nu(\Lambda)\gamma^{\pm}(\Lambda)}{G(\Lambda)};\;\;\;
\gamma^{\pm}(\Lambda)\equiv\frac{\mu\,{\cal M}^{\pm}}{2V^*}
=\left[1\mp\frac{c\sqrt{\nu(\Lambda)}}{\Lambda
G(\Lambda)}\right]^{-1};
\label{Cutof} \\
&& E^{\pm}_0(\Lambda)=\Lambda^2\left\{G(\Lambda)\left[
\frac{3}{5\nu(\Lambda)}+(1\pm2)\right]\mp\epsilon(\Lambda)\right\};
\qquad \mbox{where :}
\nonumber  \\
&&\nu(\Lambda)=\nu_0+\frac{\nu_1}{\Lambda}+
\frac{\nu_2}{\Lambda^2}+\ldots;\;\;\mbox{and the same for}\;
{\cal M}^{\pm}(\Lambda),\,\epsilon(\Lambda),\,\gamma^{\pm}(\Lambda),\,
G(\Lambda).
\nonumber
\eea
For $G_0,\nu_0\neq 0$ the finiteness of the quantities follows:
\bea
&& \gamma_0=1;\;\;\gamma^{\pm}_1=\pm c\frac{\sqrt{\nu_0}}{G_0};\;\;
{\cal M}_0=\frac{\nu_0}{2G_0};\;\;
{\cal M}^{\pm}_1=\frac{1}{2G_0}\left[\nu_1-2{\cal M}_0(G_1\mp c\sqrt{\nu_0})
\right];
\label{01}  \\
&& E^{\pm}_0=\Lambda^2\left[G_0\left(\frac{3}{5\nu_0}+(1\pm 2)\right)\mp
\epsilon_0\right]+\Lambda\left[G_1\left(\frac{3}{5\nu_0}+(1\pm 2)\right)
\mp\epsilon_1 -\frac{3G_0\nu_1}{5\nu^2_0}\right]+
\nonumber  \\
&& +\left[G_2\left(\frac{3}{5\nu_0}+(1\pm 2)\right)\mp\epsilon_2-\frac{3}{5}
\left(\frac{G_0\nu_2+G_1\nu_1}{\nu^2_0}-\frac{G_0\nu^2_1}{\nu^3_0}\right)
\right].
\label{012}
\eea
Thus, the renormalization conditions imply that first and second square
brackets in (\ref{012}) vanish. As well as in the previous case,
the bound state equation (\ref{bndst}) for J=0 serves as a transmutation
condition and looks like :
\bea
&& D^{\PP}_0\left(-iM_2(\PP)\right)=(\gamma-1)^2-\mu
{\rm J}^{\PP}_0\left(-iM_2(\PP)\right)\left[\frac{\lambda}{\mu}+
\gamma <k^2>-\PP^2-(2-\gamma)b^2\right]\Rightarrow
\nonumber  \\
&& \Rightarrow (\gamma-1)^2-\frac{\gamma}{\Lambda^3\xi}
\left(3\xi\Lambda-b\right)\left[\Lambda^2\left(\nu+
\frac{3}{5}\gamma\right)-\PP^2-(2-\gamma)b^2\right]=0;
\nonumber  \\
&& \mbox{where :}\;M_2(\PP)\equiv \E{q=ib}\Rightarrow
\frac{\PP^2}{4{\cal M}^{\pm}}+2E^{\pm}_0-\frac{b^2}{{\cal M}^{\pm}},
\qquad \mbox{and for:}
\label{DMb} \\
&& \xi=\frac{2}{3\pi};\;\;\;Y=\frac{\frac{3}{5}\gamma+\tilde{\nu}}{2-\gamma};
\;\;\;\Gamma=\frac{(\gamma-1)^2}{\gamma(2-\gamma)};\;\;\;
\tilde{\nu}(\Lambda)=\nu (\Lambda)-\frac{\PP^2}{\Lambda^2};\;\;\;
b=\Lambda z(\Lambda);\;\;\;
\nonumber  \\
&& z(\Lambda)=z_0+\frac{z_1}{\Lambda}+\ldots ,\quad
\mbox{and the same for}\;\;Y(\Lambda),\,\Gamma (\Lambda),
\label{XYGb}
\eea
it leads to cubic equation: $ z^3-3\xi z^2-Yz+\xi (3Y-\Gamma)=0 $. Now the
finiteness of $b$ implies the condition $z_0=0$ which together with
(\ref{01}), (\ref{012}) means that:
\bea
&& 3Y_0=\Gamma_0=0;\;\;\;\nu_0=-\frac{3}{5}\gamma_0=-\frac{3}{5};\;\;\;
{\cal M}_0=-\frac{0.3}{G_0};\;\;\;\epsilon_0=2G_0;\;\;
(\mbox{ so }\nu_0,\,G_0<0);
\nonumber  \\
&& \gamma^{\pm}_1=\pm i\frac{c}{G_0}\sqrt{\frac{3}{5}};\;\;\;
\epsilon_1=2G_1 \pm\frac{5}{3}G_0\nu_1;\;\;\;
Y_1=\frac{3}{5}\gamma_1+\nu_1; \quad \mbox{ if in addition,}
\label{rnrmz} \\
&&\epsilon_2=2G_2,\;\mbox{ then: }
E^+(k)\Rightarrow E^-(k)=\frac{k^2}{2{\cal M}_0}-\frac{5}{3}
\left(G_0\nu_2+G_1\nu_1+\frac{5}{3}G_0\nu^2_1\right),
\nonumber
\eea
and if besides $\nu_1=0$, then the "bare" spectrum becomes also unique.

The renormalized T-matrix for J=0 (\ref{TT0}) after subtraction
\bea
&& \DD{0}\Rightarrow \DD{0}-D^{\PP}_0\left(-iM_2(\PP)\right),\quad
\mbox{looks like :}\quad\;\;\TT{0}{k}=
\label{T0sb} \\
&&=-\frac{8\pi}{\cal M}\cdot\frac{\Lambda^2\left(\nu+\frac{3}{5}\gamma\right)
-\PP^2+q^2+k^2(1-\gamma)}{\left[\Lambda^2\left(\nu+\frac{3}{5}\gamma\right)-
\PP^2\right](b\pm iq)+(2-\gamma)\left[\frac{2}{\pi}\Lambda (b^2+q^2)-
\left(b^3\pm (iq)^3\right)\right]},
\nonumber
\eea
and under the conditions (\ref{rnrmz}) takes the form:
\be
\TT{0}{k}\mid_{\Lambda\rightarrow\infty}=-\frac{8\pi}{{\cal M}_0}\cdot
\frac{Y_1}{(Y_1+b\mp iq)(b\pm iq)}.
\label{T0rnrm}
\ee
Corresponding boundstate wave function with J=0 has the same form as for
the previous case (\ref{FB1}). Note that for the last case the Galilean
invariance which means independence on $\PP$ of both scattering and
boundstate eigenfunctions is restored manifestly only due to the applied
renormalization procedure.
For J=1 the T-matrix (\ref{TT1}) becomes:
\be
\TT{1}{k}=\frac{2\mu}{(2\pi)^3}\cdot\frac{(\ka\cdot\q)}{1-\frac{2}{3}\mu
\I{1}},
\label{T1}
\ee
and for $\gamma_0=1$ it tends to zero like $\Lambda^{-3}$ with
$\Lambda\rightarrow\infty$. So, there are no any scattering and bound
states with J=1 for such selfadjoint extension,
defined by (\ref{01}), (\ref{rnrmz}).

To find another extension let us apply the above described
renormalization procedure to the case J=1. The denominator of (\ref{T1})
with $q\Rightarrow ib_1$, $b_1=\Lambda y$, gives the following transmutation
conditions: $y^3-3\xi y^2-\xi\left(3/(2\gamma)-1\right)=0$, $y_0=0$,
which mean that $\gamma_0=3/2$. Then the finiteness of ${\cal M}_0$
leads to the requirements: $G_0=G_1=\nu_0=\nu_1=\epsilon_0=\epsilon_1=0$.
However, for this case both T-matrices $\TT{0}{k},\;\TT{1}{k}$ tend to
zero like $\Lambda^{-1}$.
So, such extension is completely equivalent to the free Hamiltinian.

The authors are grateful to A.Andrianov and A.Kaloshin for constructive
discussions.

\end{document}